\documentclass[prd,showpacs,preprintnumbers,floatfix,superscriptaddress,10pt]{revtex4}
\usepackage[mathscr]{eucal}
\usepackage{color}
\usepackage[dvips]{graphicx}
\usepackage{epsf}
\usepackage{bm}
\usepackage{epsfig}
\usepackage{feynmf}
\usepackage{amsmath}

\newcommand{\cd}{\! \cdot \!}
\newcommand{\be}{\begin{equation}}
\newcommand{\ee}{\end{equation}}
\newcommand{\ba}{\begin{eqnarray}}
\newcommand{\ea}{\end{eqnarray}}

\begin{document}

\title{Shear viscosity due to phonons in superfluid neutron stars}

\author{Cristina Manuel}
\email{cmanuel@ieec.uab.es}
\affiliation{Instituto de Ciencias del Espacio (IEEC/CSIC) Campus Universitat Aut\`onoma de Barcelona, Facultat de Ci\`encies, Torre C5, E-08193 Bellaterra (Barcelona), Catalonia, Spain}

\author{Laura Tolos}
\email{tolos@ice.csic.es}
\affiliation{Instituto de Ciencias del Espacio (IEEC/CSIC) Campus Universitat Aut\`onoma de Barcelona, Facultat de Ci\`encies, Torre C5, E-08193 Bellaterra (Barcelona), Catalonia, Spain}

\pacs{05.60-k,47.37.+q;95.30.Lz}

\begin{abstract}
{We compute the contribution of phonons to the shear viscosity $\eta$ 
in  superfluid  neutron stars, assuming  neutron pairing in a $^1S_0$ channel.
We use a Boltzmann equation amended by a collision term that takes into account the binary collisions of
phonons. We use effective field theory techniques to extract the phonon scattering rates, written as a function of
the equation of state (EoS) of the system. Our formulation is rather general, and can be used to extract the shear viscosity due
to binary collisions of  phonons for other superfluids, such as the cold Fermi gas in the unitarity limit. We find that $\eta \propto 1/T^5$,
the proportionality factor depending on the EoS of the system.
Our results indicate that the phonon contribution to $\eta$ cannot be ignored and might have relevant effects in
the dynamics of the different oscillation modes of the star.}
\end{abstract}

\date{October, the 4th 2011}
\maketitle

\section{Introduction}

After Migdal's observation that superfluidity of neutron matter may occur in the core of compact stars \cite{Migdal}, this possibility has
been seriously explored and taken into account to explain very different neutron star phenomena, such as  pulsar
glitches or the star cooling.  Superfluidity would also affect  many other aspects of the neutron star dynamics,
the corresponding hydrodynamics being essentially different from that of a normal fluid  ~\cite{landaufluids,IntroSupe}.
In particular, rotational and vibrational properties of the star, or the dynamics of the star's oscillations would be specially sensitive to
the presence of a superfluid.

 Neutron matter superfluidity  occurs at low temperatures
after the appearance of a quantum condensate, associated to neutron pairing.
The condensate spontaneously breaks the global $U(1)$ symmetry associated to baryon number conservation.
In such a case  Goldstone's theorem predicts the existence of a low energy mode that at sufficiently low momentum has  a linear dispersion law, and which is essential to explain the property of superfluidity. 
We will refer generically to this mode as superfluid phonon, or phonon for simplicity. Although we are here mainly interested in neutron matter, the appearance of  a phonon is however a generic feature that happens in all superfluid systems.

The phonons dominate the physics in the superfluid at long wavelength. At very low temperatures they also dominate the thermal corrections to the
thermodynamical and hydrodynamical properties of the superfluid. Surprisingly, even if this fact is very well-known, the presence and effects of the
 phonons have been largely ignored in the computation of all the transport coefficients that are needed
to study the dynamics of superfluid neutron star's oscillations  \cite{Flowers,Cutler,Andersson:2004aa}. We are only aware of  Ref.~\cite{Aguilera:2008ed},
which studies the thermal conductivity due to the superfluid phonons in the inner crust of the star.
 
It is our intention to compute the phonon contribution to the different transport coefficients of superfluid neutron matter. In this manuscript we start with
the computation of the phonon contribution to the shear viscosity $\eta$  in the core of superfluid neutron stars, assuming that the neutrons pair in a $^1S_0$ channel.
We leave the case of neutron pairing in other channels, the computation valid for the inner crust of the star, or the computation of other transport coefficients for future studies.

In order to compute the phonon contribution to the different transport coefficients one needs to assess the relevant phonon collisions which are responsible for 
the transport phenomena under discussion. There has been a lot of progress in this direction in recent years. 
The leading phonon interactions can be completely determined by the equation of state (EoS) of the system, 
as can be seen after writing the corresponding effective field theory at leading order in a derivative expansion \cite{Son:2002zn,Son:2005rv}.
 
 Our computation of the shear viscosity is very similar to the same sort of computations for other superfluid systems \cite{Manuel:2004iv,Rupak:2007vp,Alford:2009jm}. In this paper
 we however exploit the universal character of the effective field theory at leading order to present a very general formulation, that only depends on the EoS
 of the system.   This will allow us to obtain 
immediately the viscosity due to the binary collisions of phonons of other superfluid systems, such as, for example,  the cold Fermi  atoms in the unitarity limit
 \cite{Rupak:2007vp,Giorgini:2008zz}.
Our formulation also will allow us to improve the value of $\eta$, if a better determination  of the EoS for the nuclear matter present in the star is achieved.

This paper is structured as follows. In Sec.~\ref{phononEFT} we review the effective field theory that describes the phonon self-interactions in a superfluid, and write all
the phonon self-couplings in a model independent way, that is, without specifying the system we are considering. We use this theory to extract the scattering amplitudes
of the binary collisions of phonons in Subsec.~\ref{cross-section}. In Sec.~\ref{EOS-section} we present and review the specific equation of state for $\beta$-stable nuclear
matter that will use in our computation of the shear viscosity. In Subsec.~\ref{general-kinetic} we use the transport equation to show how the value of $\eta$ can be extracted
for a generic superfluid.  In Subsec.~\ref{numericalsec} we present numerical results for the value of $\eta$ when we consider the EoS corresponding to $\beta$-stable nuclear,
and also that of a free  neutron gas. In Subsect.~\ref{comparison} we compare our results with those obtained for $^4$He and for the cold Fermi gas in the unitarity limit.
We end up in Sec.~\ref{discu} with a discussion of our results. In all the manuscript we use natural units, except in the plots when we give numerical values of different physical
quantities.

\section{Superfluid phonon interactions}
\label{phononEFT}


In this Section we first review the effective field theory (EFT) that at leading order in a derivative expansion describe the phonon interactions. 
We parametrize all phonon self-couplings in terms of the density, speed of sound, and derivatives of the speed of sound with respect to the
density in a model independent way. In Subsect.~\ref{cross-section} we use this very general
formulation to extract the scattering amplitudes associated to the binary collisions of phonons.

\subsection{The effective field theory}

The superfluid phonon is the Goldstone mode associated to the spontaneous symmetry breaking
of a $U(1)$ symmetry, which corresponds to particle number conservation.
EFT techniques can be used to write down the effective Lagrangian associated to the superfluid phonon.  The effective Lagrangian
is then presented as an expansion in derivatives of the Goldstone field, the terms of this
expansion being 
restricted by symmetry considerations.
 The coefficients of the Lagrangian can be in principle computed from the microscopic theory, through a standard
matching procedure, and thus  they depend on the short range physics of the system under
consideration.

It has been known for a while that the Hamiltonian of the Goldstone
mode of a superfluid system is entirely fixed by  the equation of state~\cite{Popov,Greiter:1989qb}. 
In some more recent publications \cite{Son:2002zn,Son:2005rv} it has been realized that at
lowest order in a derivative expansion the Lagrangian reads
\cite{Son:2005rv}
\begin{equation}
\label{LO-Lagran}
\mathcal{L}_{\rm LO} =P (X) \ ,
\end{equation}
where $P(\mu)$ and $\mu$ are the pressure and chemical potential, respectively, of the superfluid at $T=0$, and
\begin{equation}
 X = \mu-\partial_t\varphi-\frac{({\bf \nabla}\varphi)^2}{2m} \ ,
\end{equation}
where $\varphi$ is the phonon field, and $m$ is the mass of the
particles that condense. After a Legendre
transform, one can get the associated Hamiltonian, which has the same form as the one used by Landau to obtain
the self-interactions of the phonons of $^4$He \cite{IntroSupe,Son:2005rv}.

This effective field theory has been used for the computation of different phonon effects in different superfluid systems. In particular, it has been shown ~\cite{Escobedo:2010uv}
that it allows to reproduce the leading thermal corrections to the speed of sound
at low temperature that were obtained by Andreev and Khalatnikov for superfluid $^4$He \cite{Andreev}.

It is convenient to re-express the phonon Lagrangian in a different way.
After a Taylor expansion of the pressure, and rescaling of the phonon field to have a canonically normalized kinetic term, one can write the Lagrangian for
the phonon field as
\begin{equation}
\label{comlag}
\mathcal{L}_{\rm LO}=\frac{1}{2}\left((\partial_t\phi)^2-v^2_{\rm ph}({\bf \nabla}\phi)^2\right)-g\left((\partial_t \phi)^3-3\eta_g \,\partial_t \phi({\bf \nabla}\phi)^2 \right)
+\lambda\left((\partial_t\phi)^4-\eta_{\lambda,1} (\partial_t\phi)^2({\bf \nabla}\phi)^2+\eta_{\lambda, 2}({\bf \nabla}\phi)^4\right)
+ \cdots
\end{equation}

The different  phonon self-couplings of Eq.~(\ref{comlag}) can be  expressed
as different ratios of derivatives of the pressure with respect to the chemical potential \cite{Escobedo:2010uv}. But for comparisons
with the condensed matter literature on superfluidity, and for the purposes of computation in this paper,
it turns out to be more convenient to express them in terms of the density, the speed of sound at $T=0$, and derivatives
of the speed of sound with respect to the density. 
First of all, it is easy to check that the phonon velocity is the speed of sound at $T=0$ 
\begin{equation}
\label{phspeed}
v_{\rm ph}=   \sqrt{\frac{\partial P}{\partial {\tilde \rho}} } \equiv c_s \ ,
\end{equation}
where ${\tilde \rho}$ is the mass density, related to the particle density $\rho$ as ${\tilde \rho} = m \rho$.

One then finds that the  dispersion law obtained from this Lagrangian at tree
level is exactly  $E_p = c_s p $.
 The three phonon self-coupling constants can be expressed as (see Appendix A of Ref.~\cite{Escobedo:2010uv})
\be
g=\frac{1}{6 \sqrt{m\rho} \ c_s } \left(1-2 \frac{\rho}{c_s}\frac{\partial c_s}{\partial \rho} \right) \ , \qquad
\eta_g = \frac{c_s}{6 \sqrt{m\rho}\  g }  \ ,
 \label{eq:g}
 \ee
 while the four phonon coupling constants are
 \begin{eqnarray}
\lambda &=& \frac{1}{24  \ m  \rho \ c_s^2 } \left( 1-8 \frac{\rho}{c_s} \frac{\partial c_s}{\partial \rho}+10 \frac{\rho^2}{c_s^2} \left( \frac{\partial c_s}{\partial \rho} \right)^2-2\frac{\rho^2}{c_s} \frac{\partial^2 c_s}{\partial \rho ^2}\right)  \ , \nonumber \\
\eta_{\lambda_2} &=&\frac{c_s^2}{8 \ m  \rho \ \lambda } \ , \qquad
\eta_{\lambda,1} = 2 \frac{\eta_{\lambda,2}}{\eta_g}  \ .
\label{eq:relations}
\end{eqnarray}

It is also possible to construct the Lagrangian for the superfluid phonon field at next to leading order (NLO). However, for our present purposes, we will not need it.
It is only convenient to remember that at NLO the phonon dispersion law receives corrections  that take the form
\begin{equation}
\label{displaw-NLO}
E_p = c_s p + B p^3 + \cdots 
\end{equation}
where the parameter $B$ can be computed by a matching procedure with the microscopic theory. The value of $B$ is relevant in order to see whether
the decay of a phonon into two is kinematically allowed or not. Only dispersion laws that curve upward can allow such processes.  The possibility of
having these phonon decay processes is important, as they affect the value of the  different transport coefficients of the superfluid. 
For the system of our interest, superfluid neutron matter, we have checked that $B < 0$, and thus we will only consider binary collisions of phonons
in the present manuscript.

\subsection{The scattering cross sections of binary collisions of phonons}
\label{cross-section}

\begin{figure}
\begin{center}
\hbox{\psfig{file=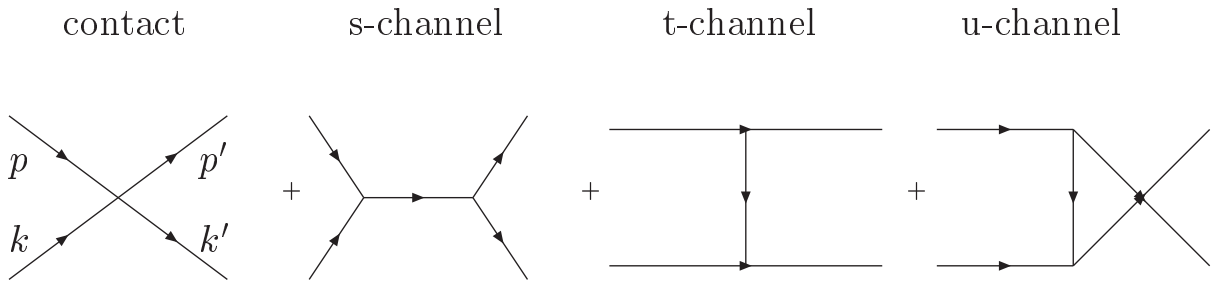,width=16cm}} \caption{
Feynman diagrams contributing to $2\leftrightarrow 2$ phonon
scattering processes.}
 \label{feyndiags}
\end{center}
\end{figure}

A necessary ingredient for the computation of the shear viscosity is the knowledge of the scattering cross section associated to
the collisions responsible for that dissipation. It is possible to extract them, at leading order in a momentum expansion, from the phonon Lagrangian 
Eq.~(\ref{comlag}). While these scattering amplitudes have been written down for some specific superfluids (that is, specifying
the value of the EoS and thus of the phonon couplings), we write them down here in a much more general form, that is, without specifying
the EoS of the system.

 The four Feynman diagrams of the binary collisions under consideration are depicted in Fig.~\ref{feyndiags}.
The contact interaction amplitude is
\ba
i {\cal M}_{c.t.} & = & - i \lambda \Big \{ 24 E_p E_k E_{p'} E_{k'}  - 4 \eta_{\lambda,1} \left( E_{p} E_{k}\  {\bf p'}\cdot {\bf k'} + E_{p} E_{p'}\ {\bf k}\cdot {\bf k'} +
E_{p} E_{k'}\ {\bf p'}\cdot {\bf k } + E_{p'} E_{k} \ {\bf p}\cdot {\bf k'} + E_{p'} E_{k'} \ {\bf p}\cdot {\bf k} \right. 
\nonumber
\\
&+& \left.   E_{k} E_{k'} {\bf p}\cdot {\bf p'} \right) + 8 \eta_{\lambda,2} \left( {\bf p}\cdot {\bf k}\  {\bf p'}\cdot {\bf k'} + {\bf p}\cdot {\bf p'} \
{\bf k}\cdot {\bf k'} + {\bf p}\cdot {\bf k'} \  {\bf p'}\cdot {\bf k}  \right) \Big \} \ ,
\ea
while the $s$-channel amplitude reads
\be
i {\cal M}_s = - 4 i g^2 G(P+K)  \left\{ E_ p K^2  + E_k P^2 + 2 (E_p+E_k) P \cdot K \right\} 
\left\{ E_{p'} K'^2  + E_{k'} P'^2 + 2 (E_{k'}+ E_{p'}) P' \cdot K' \right\} \ ,
\ee
where we used the notation $P \cdot K \equiv  E_p E_k - 3 \eta_g {\bf p} \cdot {\bf k}$, and $G$ is the phonon propagator.
The $t$- and $u$-channel amplitudes can be obtained from the $s$-channel one by using the crossing symmetry
$i {\cal M}_t =i {\cal M}_s ( K \leftrightarrow -P')$ and $i {\cal M}_u =i {\cal M}_s ( K \leftrightarrow -K')$.

As found out in Ref.~\cite{Manuel:2004iv}, if one uses the bare phonon propagator  as obtained from the Lagrangian 
Eq.~(\ref{comlag}),  one  finds collinear singularities in small angle collisions  where one virtual phonon is
exchanged. In order to avoid collinear singularities in the intermediate steps of the computation of the shear viscosity
  one has to regulate the phonon propagator. One possibility is to resum the
phonon propagator including the one-loop
thermal damping. Thus, one writes
\begin{equation}
G(P) = \frac{1}{E_p^2- c_s^2  p^2 + i \,{\rm Im}\, \Pi(P)}  \ .
\end{equation}
The imaginary part of the self-energy is extracted after a one-loop computation (see Ref.~\cite{Escobedo:2010uv}). For
 space-like momenta and $p \ll T$  \cite{Manuel:2004iv},
\begin{equation}
 {\rm Im} \,\Pi(P) = \frac{\pi^3}{15 c_s^5} \frac{T^4}{\rho} \left( 1+ \frac{\rho}{c_s}\frac{\partial c_s}{\partial \rho}\right)^2 \frac{p_0^3}{c_s p} \Theta( c_s^2 p^2 - p_0^2) \ ,
 \end{equation}
while for time-like momenta, and $p_0 > 0, p \rightarrow 0$ one finds \cite{Rupak:2007vp}
\begin{equation}
 {\rm Im} \,\Pi(P) = \frac{18 g^2}{16^2 \pi c_s^3} \left( 1 + \frac{\eta_g}{c_s^2} \right)^2 p_0^6 \frac{e^{p_0/2T} +1}{e^{p_0/2T}-1} \ .
 \end{equation}

Another possible way  of regulating the phonon propagator is to work with the EFT at next to leading order, using the non-linear dispersion law
Eq.~(\ref{displaw-NLO}). However, for the computation of the
shear viscosity in superfluid neutron matter we have checked that any of these two possible regulations give the same result.
This is due to the fact that the shear viscosity is dominated by the large angle collisions, the viscosity being 
insensitive to collinear collisions.

\section{Equation of state of superfluid neutron matter}
\label{EOS-section}


In order to obtain the speed of sound at $T=0$ as well as the different phonon self-couplings one needs to determine the EoS for neutron matter in neutron stars.
Then one can get the speed of sound, and all the derivatives of the speed of sound with respect to the density.
A common benchmark for a nucleonic equation of state is the one obtained by  Akmal,  Pandharipande and  Ravenhall \cite{ak-pan-rav}
(APR for short).
 These authors studied neutron matter as well as symmetric and $\beta$-stable nuclear matter. Later on  Heiselberg and Hjorth-Jensen \cite{hei-hjo} parametrized the APR EoS of nuclear matter in a simple form, which will subsequently  be used in this manuscript. It should be remarked that the effect of neutron pairing is not considered in the previous references. However, because neutron pairing only affects
those neutrons which are closed to their Fermi surface, one does not expect that the effect of pairing might have a big impact in the EoS. For this reason, we will neglect the effect, and consider that it only
represents a  small correction.

The binding energy per nucleon (E/A) in nuclear matter consists of a compressional term and a symmetry term:
\begin{eqnarray}
E/A=E_{comp}(\rho)+S(\rho)(1-2 x_p)^2=\mathcal{E}_0 u \frac{u-2-\delta}{1+\delta u}+S_0 u^{\gamma}(1-2x_p)^2.
\label{eq1}
\end{eqnarray}
Here $u$ is the ratio of the nucleon density ($\rho$) to nuclear saturation density ($\rho_0=0.16  \ {\rm fm^{-3}}$),  $u=\rho/\rho_0$, and $x_p=\rho_p/\rho_0$ is the proton fraction. The nucleon density is given by
\begin{eqnarray}
\rho=\nu \int^{p_{F}}_0 \frac{d^3p}{(2 \pi)^3} \ ,
\end{eqnarray}
with $p_F$ being the Fermi momentum and $\nu$ the degeneracy factor. In nuclear matter $\nu$ is 4. The compressional term is parametrized by a simple form which reproduces the saturation density, binding energy and compressibility. The binding energy per nucleon at saturation density excluding Coulomb energies is $\mathcal{E}_0=15.8 \ {\rm MeV}$ and the parameter $\delta=0.2$ was determined by fitting the energy per nucleon at high density to the EoS of Akmal \cite{ak-pan-rav} with three-body forces and boost corrections, but taking the corrected values from Table 6 of \cite{ak-pan-rav}. The corresponding compressibility is $K_0=18 \ \mathcal{E}_0/(1+\delta) \simeq 200 \ {\rm MeV}$, within the experimental values. For the symmetry term Heiselberg et al. obtained $S_0=32 \ {\rm MeV}$ and $\gamma=0.6$ for the best fit.

The EoS is then
\begin{equation}
\mathcal{E} (\rho,x_p)=(m+E/A (\rho,x_p)) \ ,
\label{eq2}
\end{equation}
with $m$ being the mass of the nucleon, while the corresponding nucleonic energy density is 
\begin{eqnarray}
\epsilon_N(\rho,x_p)=\mathcal{E}(\rho,x_p) \rho \ .
\label{eq3}
\end{eqnarray}

For neutron star matter made of neutrons, protons and electrons, the total energy density is the sum of the nucleonic  contribution (neutrons and protons), $\epsilon_N$, and the one for electrons, $\epsilon_e$,
\begin{eqnarray}
\epsilon(\rho,x_p,\rho_e)= \epsilon_N(\rho,x_p)+\epsilon_e(\rho_e) .
\end{eqnarray}
The pressure includes also both contributions
\begin{eqnarray}
P(\rho,x_p,\rho_e)=P_N(\rho,x_p)+P_e(\rho_e) \ ,
\end{eqnarray}
where the nucleonic and electronic contributions to the pressure are given by
\begin{eqnarray}
P_N(\rho,x_p)&=&\mu_n (\rho,x_p) \ (1-x_p) \ \rho + \mu_p (\rho,x_p) \ x_p \rho-\epsilon_N(\rho,x_p) , \nonumber \\
P_e(\rho_e)&=& \mu_e(\rho_e) \rho_e-\epsilon_e(\rho_e)  \ ,
\end{eqnarray}
being $\mu_i$ the chemical potential of each specie.
Those chemical potentials are calculated as
\begin{eqnarray}
\mu_n(\rho,x_p)&=&\frac{\partial \epsilon_N(\rho,x_p)}{\partial \rho_n}=\frac{\partial \epsilon_N(\rho,x_p)}{\partial \rho}-\frac{\partial \epsilon_N(\rho,x_p)}{\partial x_p} \frac{x_p}{\rho} \ , \nonumber \\
\mu_p(\rho,x_p)&=&\frac{\partial \epsilon_N(\rho,x_p)}{\partial \rho_p}=\frac{\partial \epsilon_N(\rho,x_p)}{\partial \rho}+\frac{\partial \epsilon_N(\rho,x_p)}{\partial x_p} \frac{(1-x_p)}{\rho} \ ,\nonumber \\
\mu_e(\rho_e)&=&\sqrt{p_{F_e}^2+m_e^2} \sim p_{F_e} \sim (3 \pi^2 \rho_e)^{1/3} .
\label{eq:mus}
\end{eqnarray}

Moreover, matter in neutron stars is thought to be in  $\beta$-equilibrium against weak decay processes among the different species. The constraints imposed  by chemical equilibrium and charge neutrality for matter made of neutrons, protons and electrons are
\begin{eqnarray}
\mu_n&=&\mu_p+\mu_e  \ ,\nonumber \\
\rho_p&=&\rho_e  \ .
\end{eqnarray}

Using the definitions for the chemical potentials of Eq.~(\ref{eq:mus}) and the $\beta$-equilibrium condition, one finds that
\begin{eqnarray}
\mu_e=\mu_n-\mu_p=-\frac{1}{\rho} \frac{\partial \epsilon_N(\rho,x_p)}{\partial x_p}=4 S_0 u^{\gamma} (1-2 x_p) \ .
\end{eqnarray}
Then, one can use the charge neutrality condition for ultrarrelativistic electrons:
\begin{eqnarray}
\mu_e=E_{F_e}=p_{F_e}=(3 \pi^2 \rho_e)^{1/3} =(3 \pi^2 \rho_p)^{1/3}  \ .
\end{eqnarray}
Using that $\rho_p=x_p \rho$, we get
\begin{eqnarray}
x_p=\frac{(4 S_0 u^{\gamma} (1-2 x_p))^3}{3 \pi^2 \ \rho } \ .
\label{eq:frac}
\end{eqnarray}
From this relation, we can find an analytic equation for the proton fraction.
Defining 
\begin{eqnarray}
a=2(4 S_0 u^{\gamma})^3/(\pi^2 \rho) \ ,
\end{eqnarray}
Eq.~(\ref{eq:frac}) reduces to
\begin{eqnarray}
a x^3+ 3 x -3 =0 \ ,
\label{eq:ec}
\end{eqnarray}
with $x=1-2x_p$. 
This equation has one real solution given by
\begin{eqnarray}
x=(-2 \ 2^{1/3} a + (6 a^2 + 2 \sqrt{a^3 (4 + 9 a)})^{
 2/3})/(2 a \ (3 a^2 + \sqrt{a^3 (4 + 9 a)})^{1/3}) ,
\end{eqnarray}
which sets the relation between the proton fraction and the total nucleonic density.

In our calculations for the shear viscosity we will use the APR EoS of  $\beta$-stable nuclear matter, in the parametrization just presented.
We have also compared our results with those  using the APR  EoS of interacting neutron matter and the EoS of a free neutron gas, that is, assuming that the interactions are very weak and their effect is neglected in the EoS.  We have checked that the scenario of only interacting neutrons leads to  very similar results for the speed of sound and all phonon self-couplings than that of $\beta$-stable nuclear matter. Those results were, however, expected due to the low proportion of protons in $\beta$-stable matter. On the other hand, the EoS of the free neutron gas shows a different behavior to $\beta$-stable matter,
and thus of the speed of sound and phonon self-couplings, and ultimately, of the shear viscosity, as we will show in the following Section.

\begin{figure}[t]
\begin{center}
\includegraphics[width=0.5\textwidth]{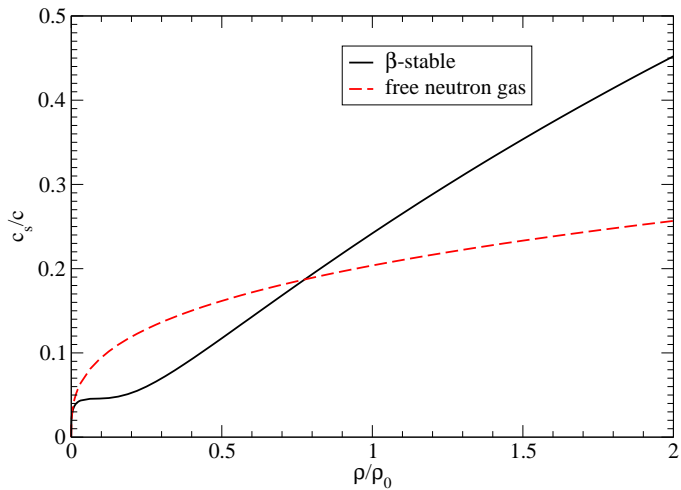}
\caption{ Ratio of the speed of sound with respect to the speed of light  $c$ obtained for a free gas of neutrons and for $\beta$-stable nuclear matter as a function of the density.}
\label{speeds}
\end{center}
\end{figure}

 In Fig.~\ref{speeds}
we plot the ratio of the speed of sound $c_s$ with respect to the speed of light $c$ for both a free gas of neutrons and for  $\beta$-stable nuclear matter as a function of the density. We observe that the assumption that the APR model for $\beta$-stable nuclear matter is non-relativistic breaks down at densities of the order of 1.5-2 $\rho_0$. For those densities, relativistic effects appear as the APR EoS includes not only two nucleon but also three nucleon interactions. This seems to suggest that at very high densities we should employ the relativistic 
version of the phonon effective field theory   \cite{Son:2002zn}. This would imply very minor changes in the computation of $\eta$ (see Ref.~\cite{Manuel:2004iv} for the treatment of a  relativistic superfluid).

In the following, we will  perform the following approximations for the computation of the  speed of sound and all three and four phonon self-couplings. First, to compute the mass
density we only take into account the nucleonic part,  as $m \gg m_e$. Further, in  $\beta$-equilibrated matter, $\rho \approx \rho_n$.  Thus, we will assume that the speed of sound can be computed as
\begin{eqnarray}
c_s(\rho,x_p) \approx \sqrt{\frac{1}{m}\frac{\partial P_N(\rho,x_p)}{\partial \rho_n}} , 
\end{eqnarray}
and, for the phonon self-couplings, we will substitute  $\rho \rightarrow \rho_n$ in Eqs.(\ref{eq:g},\ref{eq:relations}).

\section{  Shear viscosity due to the binary collisions of  superfluid phonons}


We report here all the computations required to extract the shear viscosity due to the binary collisions of phonons in a non-relativistic
superfluid. Our final goal is to extract the value of the viscosity in $\beta$-stable neutron matter, using the EoS just presented in the
previous Section. Because to leading order in a momentum expansion the main interactions leading to the value of the viscosity
are set by the EoS of the system, we first present in Subsect.~\ref{general-kinetic} a very general formulation of the computation.
 We follow here the same methodology used for the study of some specific superfluids \cite{Manuel:2004iv,Rupak:2007vp,Alford:2009jm}, but we are aiming here to have a broad and general formulation that
can be used for different  superfluid systems, or for future improvements of the knowledge of the EoS of the superfluid neutron matter contained in the star. 

In Subsect.~\ref{numericalsec} we present the numerical results of the shear viscosity when the APS EoS 
discussed in the previous
Section is chosen. To see the effect of the nucleon interactions, we also compute the viscosity with the EoS of a free neutron gas.
In Subsect.~\ref{comparison} we compare our results with those obtained for $^4$He and for the cold Fermi gas in the unitarity limit.


\subsection{General formulation}
\label{general-kinetic}


The shear viscosity $\eta$ enters as a dissipative term in the energy-momentum tensor $T_{ij}$. For small deviations from equilibrium one
finds
\be
\label{shear-stress}
 \delta T_{ij}=- \eta \tilde V_{ij}  \equiv  - \eta\left( \partial_i V_j+ \partial_j V_i -\frac 23 \delta_{ij} \nabla \cd {\bf V} \right) \ , 
  \ee
where ${\bf V}$ is the three fluid  velocity of the normal component of the system.

The superfluid phonon contribution to  the energy-momentum
of the system can be expressed as
 \be
 T_{ij}= c_s^2 \int \frac{d^3 p}{(2 \pi)^3}  \frac{ p_i p_j}{E_p} f(p,x)  \ , 
  \ee
where $f$ is the phonon distribution function. The distribution function obeys the Boltzmann equation \cite{IntroSupe}
 \be
  \label{transport}
   \frac{df}{dt} = \frac{\partial
f}{\partial t}+ \frac{\partial E_p}{\partial \bf p} \cdot \nabla f= C[f] \ ,
\ee
where we have assumed to be in the superfluid rest frame, and
$C[f]$ is the collision term. For the computation of the shear viscosity it is enough to consider binary collisions, and thus
\be
C[f] = \frac {1}{2E_p} \int \frac{d^3 k}{(2 \pi)^3 2E_k} \frac{d^3 p'}{(2 \pi)^3 2E_{p'}} \frac{d^3 k'}{(2 \pi)^3 2E_{k'}} (2\pi)^4 \delta^{(4)}( P+K-P'-K') \frac 12 |{\cal M}|^2 D \ ,
\ee
where
\be
D = f(p') f(k') (1+f(p))(1+f(k)) - f(p) f(k) (1+f(p'))(1+f(k')) \ ,
\ee
and ${\cal M}$ is the $2 \leftrightarrow 2$ scattering matrix
\begin{equation}
{\cal M} = {\cal M}_{c.t.} +  {\cal M}_s + {\cal M}_t + {\cal M}_u \ .
\end{equation}
The explicit expressions for the  contact interaction, and $s$-, $t$- and $u$- channel amplitudes were 
given in Subsec.~\ref{cross-section}.

This collision term has the property that it vanishes 
when evaluated using 
 the phonon equilibrium distribution function 
\be
f_{\rm eq} = \frac{1}{e^{E_p/T} -1}  \ ,
\ee
so that $C[f_{\rm eq}]=0$.

  For the computation of the transport
coefficients we have to consider small departures from equilibrium
so that the distribution function can be written as
 \be
  f= f_{\rm eq}+
\delta f \ ,
 \ee 
and linearize the transport equation in $\delta f$. For the computations of the shear
viscosity, one assumes a particular dependence of $\delta f$. More particularly 
\be
\delta f = - h(p) p_{kl} V_{kl} \frac{f_{\rm eq} ( 1+ f_{\rm eq})}{T}    \ ,
\ee
where $h(p)$ is an unknown function, $V_{kl}$ is the tensor defined in Eq.~(\ref{shear-stress}), and we have defined the tensor
\be
p_{kl} = p_k p_l - \frac 13 \delta_{kl} p^2 \ .
\ee

One can extract the value
of  the shear viscosity 
computing the contribution that this distribution function has in the energy-momentum of the system. Thus, one finds
 \be
 \label{1stexpsh}
 \eta=\frac{4 c_s^2 }{15 T }\int
\frac{d^3p}{(2\pi)^3} \frac{p^4}{2 E_p}  f_{\rm eq}(1+f_{\rm eq}) h(p) \ .
 \ee

To obtain the shear viscosity we have to know the value of $h(p)$. For this purpose we linearize the transport equation Eq.~(\ref{transport}).
The linearized  collision term reads
\ba
\delta C &=&
 \frac {1}{2E_p T}  \int \frac{d^3 k}{(2 \pi)^3 2E_k} \frac{d^3 p'}{(2 \pi)^3 2E_{p'}} \frac{d^3 k'}{(2 \pi)^3 2E_{k'}} (2\pi)^4 \delta^{(4)}( P+K-P'-K') \frac 12 |{\cal M}|^2 \nonumber \\
 & \times &
 f^{\rm eq}_p f^{\rm eq}_k ( 1+f^{\rm eq}_{p'})(1+  f^{\rm eq}_{k'}) \left( h(p) p_{ij} + h(k) k_{ij} - h(p') p'_{ij} - h(k') k'_{ij} \right)  V_{ij} \nonumber \\
 & \equiv & \frac {1}{2E_pT} F_{ij}[h(p)] V_{ij} \ ,
   \ea
so that the linearized Boltzmann equation reads
\be
c_s \frac{ f_{\rm eq}(1+f_{\rm eq})  }{2pT} p_{ij} V_{ij} = \frac {1}{2E_pT} F_{ij}[h(p)] V_{ij} \ .
\ee

With this last result one realizes that the shear viscosity can also be written as
\be
\label{2onexpsh}
 \eta=\frac{2 }{5 T} \int
\frac{d^3p}{(2\pi)^3} p_{ij} h(p) F_{ij}[h(p)] \ .
 \ee

It is possible to solve the transport equation using variational methods.
We will follow here the same strategy as that followed in Refs.~\cite{Rupak:2007vp,Alford:2009jm}, and
select the trial functions as
\be
\label{trialfunction}
 h(p)= p^n \sum_{s=0}^{\infty} b_s B_s(p) 
 \ ,\ee
 where $n$ is a parameter that will be determined in a variational procedure, and $B_s(p)$ are 
  orthogonal polynomials of order $s$ and are defined such that the coefficient of the
  highest power $p^s$ is 1, and the orthogonality condition
  \be
  \int \frac{d^3p}{(2\pi)^3} \frac{   f_{\rm eq}(1+f_{\rm eq}) }{2 E_p} p_{ij} p_{ij} p^n B_r(p) B_s(p) =A_{r}^{(n)} \delta_{rs}
    \ee 
 is satisfied.
 
 Using this form of the solution one can check that Eq.~(\ref{1stexpsh}) gives
 \be
 \label{shear1}
 \eta= \frac{2 c_s^2}{5T} b_0 A_0^{(n)}  \ , \ee 
  where
 \be
 A_0^{(n)} = \frac{T^{6+n}}{6 \pi^2 c_s^{7+n}} \Gamma(6+n) \zeta(5+n)  \  ,
   \ee
 where $\Gamma(z)$ and $\zeta(z)$ stand for  the Gamma and Riemann zeta functions, respectively.
 
On the other hand, the expression of the shear viscosity given by Eq.~(\ref{2onexpsh}) reads
 \be
 \label{shear2}
 \eta = \sum_{s,t=0}^{N=\infty} b_s b_t M_{st}  \ ,
  \ee
 where
 \be
 M_{st} = \frac {1}{10T}   \int_{p,k,p',k'} 
  (2\pi)^4 \delta^{(4)}( P+K-P'-K') \frac 12 |{\cal M}|^2 
 f^{\rm eq}_p f^{\rm eq}_k ( 1+f^{\rm eq}_{p'})(1+  f^{\rm eq}_{k'}) \Delta_{ij}^s \Delta_{ij}^t  \ ,
  \label{eq:mst}
  \ee
 and
 \be
 \Delta_{ij}^t = B_t(p) p^n p_{ij} + B_t(k) k^n k_{ij} - B_t(p') p'^n p'_{ij} - B_t(k') k'^n k'_{ij}  \ ,
    \ee
where we have defined the shorthand notation 
\be
\int_p \equiv \int  \frac{d^3 p}{(2 \pi)^3 2E_{p}} \ .
\ee

Requiring that the two forms of the shear viscosity,  Eqs.~(\ref{shear1}) and (\ref{shear2}),  be equal implies
\be
b_0 = \frac{2 c_s^2}{5T} A_0^{(n)} ( M^{-1})_{00}  \ ,
\ee
so that
\be
\eta = \frac{4 c_s^4}{25 T^2} (A_0^{(n)})^2 ( M^{-1})_{00} \ .
\ee

In practical terms one performs the study by limiting the number of orthogonal polynomials included
in the analysis. One can prove that 
\be
\eta \geq  \frac{4 c_s^4}{25 T^2} (A_0^{(n)})^2 ( M^{-1})_{00}  \ ,
\ee
for a particular value of $n$ and the number of orthogonal polynomials considered in the study.

The evaluation of Eq.~(\ref{eq:mst}) requires the calculation of a
 five-dimensional integral. This can be realized after  eliminating from Eq.~(\ref{eq:mst}) the $\vec{p} \ '$ integral using the  $\delta$-function of 
 momentum conservation. The  $\delta$-function of energy conservation eliminates the integral over the magnitude of $\vec{k} \ '$. Then, we are left with a eight-dimensional integral that can be further simplified selecting the $z$ axis along the vector $\vec{p}$ and realizing that only the difference in the two remaining azimuthal angles matters. Therefore, the problem is reduced to a five-dimensional integral  over the magnitudes $p$ and $k$, two polar angles corresponding to $k$ and $k'$, and one azimuthal angle  (see Appendix C of Ref.~\cite{Alford:2009jm}, or
 \cite{Manuel:2004iv} ). This calculation can be performed numerically using the $Vegas$ Monte Carlo algorithm \cite{lepage1}.

After a very simple dimensional analysis, it is very easy to check that the shear viscosity due to the binary collisions of superfluid phonons scales with the temperature
as  $1/T^5$, as it occurs in other superfluid systems.

A variational numerical analysis can be performed once one specifies the choice of the EoS, and has the proper values of all the phonon self-couplings.
However, independently of the specific choice of the EoS, one gets the same form of the solution for different superfluid systems \cite{Manuel:2004iv,Rupak:2007vp,Alford:2009jm}. One finds that the value of
  $\eta$ is maximized for the choice $n=-1$, and for the lowest number of
polynomials one can take in the expansion (\ref{trialfunction}).  We will thus use this form of the solution to get the numerical values of the shear viscosity
in $\beta$-stable nuclear matter.


\subsection{Numerical values for the viscosity in superfluid neutron matter}
\label{numericalsec}


\begin{figure}[t]
\begin{center}
\includegraphics[width=0.5\textwidth]{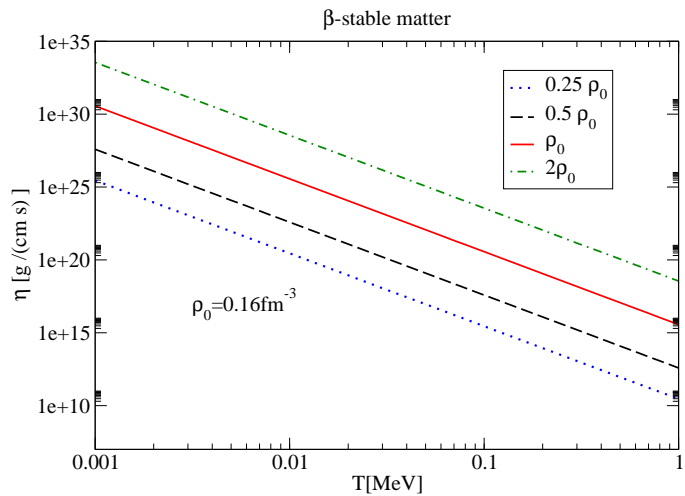}
\caption{ Shear viscosity of $\beta$-stable nuclear matter as function of temperature for four diferent densities }
\label{fig:sheartemp}
\end{center}
\end{figure}

In this Subsection we present numerical results for the shear viscosity in $\beta$-stable nuclear matter, using the APR EoS discussed in Sec.~\ref{EOS-section}.
 We have not been able to find
a closed analytical expression for the value of $\eta$, as the scattering cross section depends on all the phonon self-couplings $g, \eta_g, \lambda,\eta_{\lambda,1}$ and $\eta_{\lambda,2}$,
and all diagrams in Fig.~\ref{feyndiags} contribute to the value of $\eta$. 

We will present numerical results for values of the density higher and lower than $0.5 \, \rho_0$, where $\rho_0=0.16 \ \rm{fm}^{-3} \approx 2.8\, 10^{14} \ \rm{g/(cm \ s)}$ is the nuclear saturation density.
It should be noted here that densities below $0.5 \, \rho_0$ correspond to the densities in the crust, while densities above that value would correspond to those of the core of the star. Our computation for the inner
crust is valid for the superfluid phonon contribution to $\eta$, but one should be aware that in the inner crust there are other light degrees of freedom that might have important contributions to the
viscosity as  well \cite{Cirigliano:2011tj}. We leave a complete study of $\eta$  in the inner  crust for  future studies.

In Fig.~\ref{fig:sheartemp} we display a logarithmic plot of the shear viscosity where the expected $1/T^5$ dependence is clearly detected.  There we present the results for
four different values of the density: 0.25 $\rho_0$, 0.5 $ \rho_0$, $\rho_0$ and $2 \rho_0$.  It is also clearly seen that  $\eta$ increases very rapidly when the  density
is increased.

Naively one could think that $\eta \propto \rho^2$, tracing the direct dependence of the phonon self-couplings $g$ and $\lambda$  on $\rho$. However, these couplings also
 depend on  powers of the speed of sound and its derivatives, which are density dependent quantities, and whose values are different for different choices of the EoS. Therefore
 the $\rho$ dependence of $\eta$ is not universal for all superfluids, contrary to what happens for the temperature dependence.

\begin{figure}[t]
\begin{center}
\includegraphics[width=0.5\textwidth]{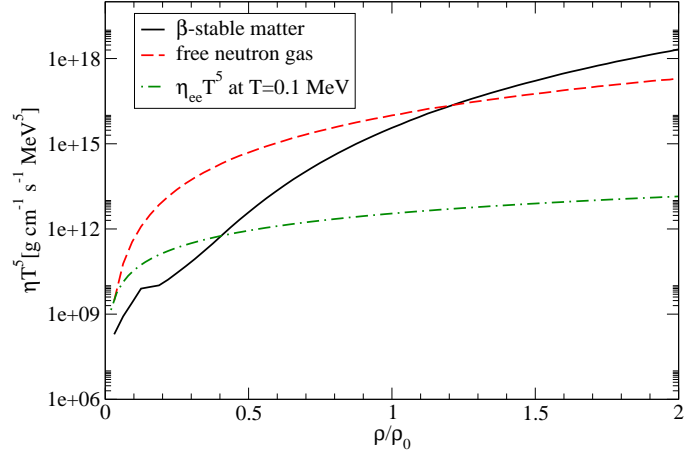}
\caption{ Shear viscosity multiplied by $T^5$ ($\eta \ T^5$) as function of density for a free neutron gas and $\beta$-stable nuclear matter. We also plot $\eta \ T^5$  due to electron-electron scattering, according to the estimate of Cutler and Lindblom, see Ref.~\cite{Cutler}. As the last is a $T$-dependent quantity, we chose $T=0.1$ MeV in that curve.}
\label{fig:sheart5}
\end{center}
\end{figure}

By exploiting that  $\eta T^5 $ is a  temperature-independent quantity, we study the $\rho$ dependence of $\eta$ in Fig.~\ref{fig:sheart5}. We consider both the APR EoS that
describes $\beta$-stable nuclear matter and  the EoS of a free  neutron gas.  Amazingly, one observes that the viscosity increases by almost three orders of magnitude when
going from $\rho_0$ to  2 $\rho_0$ for $\beta$-stable matter, the increase being much more moderate for a free neutron gas. This fact is related to the behavior of $c_s$ seen in Fig.~\ref{speeds}.  Our plot in  Fig.~\ref{fig:sheart5}  teaches us that the different choices of EoS  have a very clear impact in the numerical values of  $\eta$.

\begin{figure}[t]
\begin{center}
\includegraphics[width=0.5\textwidth]{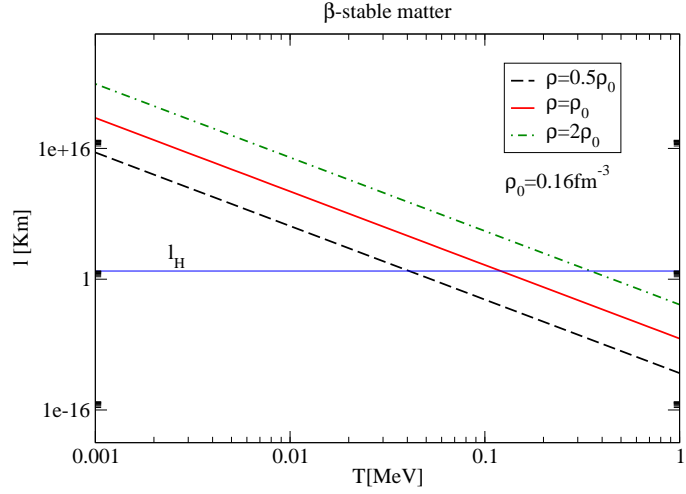}
\caption{ Phonon mean-free path in $\beta$-stable matter as function of temperature for different densities. The straight line indicates the hydrodynamic limit: $l_H < 10 \ {\rm Km}$.}
\label{fig:mfp}
\end{center}
\end{figure}

A relevant discussion also is to know the temperature regime  where our results are valid. The density of superfluid phonons becomes very dilute at very low $T$, and then it might
be difficult to maintain a hydrodynamical description of their behavior. A relevant parameter that measures the validity of  hydrodynamics is the mean free path. Hydrodynamics is only valid when the mean free path is 
smaller that the typical macroscopic length of the system, in this case, the radius of the star.
We show in Fig.~\ref{fig:mfp} the mean free path of phonons within $\beta$-stable matter for $0.5 \rho_0$, $\rho_0$ and $2 \rho_0$ as a function of temperature. We also indicate an estimate of the limit of 10 Km, the radius of the star,  above which phonon hydrodynamics should  become invalid in neutron stars.  The mean free path was extracted from the computation of $\eta$ as
\cite{Alford:2009jm}
\begin{equation}
l=\frac{\eta}{n <p>} \ ,
\end{equation}
where $<p>$ is the thermal average momentum, and $n$ is the phonon density:
\begin{equation}
<p>=2.7 \frac{T}{c_s} \ , \qquad
n=\int \frac{d^3p}{(2\pi)^3} f_p=\xi(3) \frac{T^3}{\pi^2 c_s^3} .
\end{equation}

The plots in  Fig.~\ref{fig:mfp}  indicate that for  temperatures below $T\sim 0.1$ MeV, the
phonon mean free path is bigger than the size of the star. This would seem to indicate that
for $T < 0.1$ MeV it is questionable to have a hydrodynamical description of the phonons of the star.
We note that the critical temperature for the phase transition to the normal phase is $T_c \sim 1$ MeV, although
this value, as well as the values we obtained for the phonon mean free path, are model dependent, that is,
they depend on how one models the nucleonic interactions and the EoS of the system. We should also warn the reader that we have only made  a tentative estimate,  as 
the phonons also collide with other particles in the star, such as the electrons. To assess the mean free path of the phonons due to their collision with the
electrons, one should derive an effective field theory describing their interactions. We leave this computation for future studies.

\subsection{Comparison with the shear viscosity due to phonons of other superfluid systems}
\label{comparison}


Having set a general formulation for the computation of the shear viscosity due to the binary collisions of superfluid phonons, we would
like to compare here our results with those obtained in the literature for other superfluid systems.

 Landau and Khalatnikov were the first to compute the viscosity due to the phonons in the regime where they were supposed to dominate
 the transport phenomena in the cold regime of   $^4$He  \cite{Landau}. Landau postulated an ad hoc Hamiltonian, valid for $^4$He. 
 The  formalism  we have followed to model the phonon interactions is very different than the one used for  $^4$He, 
 but as shown in Ref.~\cite{Son:2005rv}, after a Legendre transformation one can convert the Lagrangian Eq.~(\ref{LO-Lagran})
 into the Hamiltonian  used by Landau and Khalatnikov in their computations. Thus, one would expect to have very similar results for the viscosity.
 For  $T < 0.8$ K, when the phonons dominate, Landau and Khalatnikov obtained \cite{Landau}
 \begin{equation}
\eta = \left(\frac{4 \pi^2}{15}\right)^2 \rho^2 c_s^2 \frac{5 \cdot 2^{17}}{3 \cdot13!  \left(1+ \frac{\rho}{c_s}\frac{\partial c_s}{\partial \rho} \right)^2} \frac{c_s^5}{T^5} \ .
\end{equation}
However, this result was obtained after making several approximations. More particularly, after considering that the fourth order term in a perturbative expansion
of the Hamiltonian is negligible (in our language, the contact interaction diagram), and considering only collisions where the momentum of one of the particle that collides
is much smaller than the momentum of the other one. While the scaling behavior with the temperature $T$ is the same as the one we found out, the numerical prefactor is not. 
The reason is due to the  approximations that 
Landau and Khalatnikov performed to obtain the scattering cross section. In fact, our numerical analysis suggests that in our superfluid  all the different Feynman diagrams contribute to $\eta$,  and that also  not only collisions where $p \ll k$ are relevant in the computation  of $\eta$.

With our computations we can easily get the value of the shear viscosity for the ultracold Fermi gas in the unitarity limit  \cite{Giorgini:2008zz}. The EoS for that system is \cite{Ho:2004zza}
\be
P = \frac{4 \sqrt{2} m^{3/2}}{15 \pi^2 \xi^{3/2}} \mu^{5/2} \,, \qquad 
\ee
where  $\xi$ is the so-called Bertsch parameter, which relates the chemical potential and the Fermi energy $\mu = \xi E_F$. One then easily finds 
the value of $\rho$ and also that
\be
 c_s^2 = \frac{2 \mu}{3 m} \ , \qquad
\frac{\rho}{c_s}\frac{\partial c_s}{\partial\rho}=\frac{1}{3} \ , \qquad 
\frac{\rho^2}{c_s}\frac{\partial^2 c_s}{\partial\rho^2}= -\frac{2}{9}  \ .
\end{equation}
The computation of the shear viscosity for the superfluid phase was performed by Rupak and Schafer. They obtained \cite{Rupak:2007vp,Schafer:2009dj}
\be
\eta = 9.3 \cdot 10^{-6} \frac{\xi^{5}}{c_s^3}\frac{T_F^8}{T^5}  \ ,
\ee
where $T_F = \mu/\xi$. We reproduce the same scaling behavior, but find a small  discrepancy in the numerical pre-factor:
\be
\eta = 3.4 \cdot 10^{-6} \frac{\xi^{5}}{c_s^3}\frac{T_F^8}{T^5}  \ .
\ee
Because the computations in Ref.~\cite{Rupak:2007vp} and ours are essentially the same, the numerical disagreement seems to be related to a different pre-factor of the scattering cross section and the imaginary part of the phonon self-energy.


\section{Discussion}
\label{discu}

We have set a formulation of the binary collisions of the phonons and compute the associated shear viscosity  of a non-relativistic superfluid in terms of 
the equation of state of the system. Our computations are rather  general, and can be used for different superfluid systems, provided they share the same
underlying symmetries \footnote{The  cold Fermi gas in the unitary limit is conformal invariant, while nuclear matter is not. However, conformal invariance only affects the power scaling of the phonon self-couplings.}.
We have exploited the strength and universal character of the effective field theory that describes the phonon interactions at leading order in a momentum and
energy expansion. In this way, we see that the binary collisions of phonons produce a shear viscosity that scales with the temperature as $\eta \propto 1/T^5$, and this
is universal for all superfluids sharing the same symmetries. The density dependence of $\eta$ is not universal, and depends on the EoS of the system under consideration.

Our primary goal in this manuscript was computing   the phonon contribution to the shear viscosity in  superfluid neutron stars, if  the 
neutrons pair in a $^1S_0$ channel. For the superfluid in the inner crust and for neutron pairing in a $^3P_2$ channel, other contributions to $\eta$ are as well important
and will be studied somewhere else. 

We believe that the results we obtained here should be considered in the study of the behavior of different oscillation modes of superfluid neutron stars.
While in the normal phase the viscosity is dominated by neutron-neutron scattering, in the superfluid phase it was considered that the viscosity should be dominated by
electron-electron scattering. 
 There are several estimates of the electron contribution to the shear viscosity (see Fig. 1 of Ref.~\cite{Andersson:2004aa} for a comparison). We note that the phonon contribution to $\eta$ is several orders of magnitude higher
than the electron contribution in a certain low temperature range (see Fig.~\ref{fig:sheart5} for an explicit comparison of $T^5 \eta$ at $T =0.1$ MeV) 
, but this statement also depends on how deep inside one looks in the star, the phonon viscosity being much larger when the density increases.
It should be also  important to assess when phonons are in a ballistic regime or not. In any case,  our results indicate that the effect of  phonons in the value of the transport
coefficients cannot be ignored, and that one should check their impact in the study of the oscillation modes of a superfluid neutron star.

 \acknowledgments{We are specially grateful to F. Llanes-Estrada for very detailed conversations on our computations. We also thank M. Braby, M. Mannarelli, A. Rios and T. Sch\"afer for useful discussions. This research was supported by Ministerio de Ciencia e Innovaci\'on under contract FPA2010-16963. LT acknowledges support from the Ramon y Cajal Research
Programme from Ministerio de Ciencia e Innovaci\'on.}

\appendix

\end{document}